\documentclass[aps,superscriptaddress,twocolumn,showpacs,preprintnumbers,amsmath,amssymb,showkeys]{revtex4}
\usepackage[cp1251]{inputenc}
\usepackage[english]{babel}
\usepackage{amssymb,amsmath}
\usepackage{amstext} 
\usepackage[dvips]{hyperref}
\usepackage[mathscr]{eucal}
\usepackage{longtable}
\usepackage{graphicx}
\begin{document}
\title{Coherent Quantum Optics Phenomena in Carbon Low-Dimensional Systems}
\author{Alla Dovlatova (a), Dmitri Yerchuck (b)\\
\textit{(a) - M.V.Lomonosov Moscow State University, Moscow, 119899, \\ (b) Heat-Mass Transfer Institute of National Academy of Sciences of RB, Brovka Str.15, Minsk, 220072}}
\date{\today}%
\begin{abstract} Brief review of the theoretical and experimental results, based mainly on the works of authors, in the application of quantum field theory  to the study of carbon low-dimensional systems - quasi-1D carbon nanotubes, carbynes and graphene with  emphasis on formation of longlived coherent states of joint photon-electron and joint resonance phonon-electron systems of given materials is presented. \end{abstract}  
\maketitle                         
\section{Introduction}
The use of optical and radio spectroscopy  methods to create coherent states in solid materials has many potential
applications, ranging from  nonlinear optics to 
solid-state quantum computing. The interest in optical or microwave methods of coherent state formation lies in existing at present  many studies  in the fact, that the coherent states of point centers in crystals  can be
efficiently excited and manipulated using optical laser fields or almost  monochromatic microwave field since there are point centers in crystals (for example, substitual phosphor donor atoms in Si and N-V centers in diamond single crystals), which  are  relatively weakly coupled to the surrounding host lattice atoms and hence
have the long coherence lifetimes needed for instance, for various optoelectronic and spintronic  devices and for elaboration of quantum computing. 

It seems to be very substantial for practical applications and even necessary the development of the  theory, which allows to predict the appropriate electronic systems and the conditions for the formation of coherent states with more long life time. Subsequent progress  in given field seems to be connected with the elaboration of theoretical models based on quantum field theory (QFT) including quantum electrodynamics (QED). Really, quantum field theory, in fact,  becomes to be working instrument in spectroscopy studies and industrial spectroscopy control. Moreover, we will show in given report, that quantized electromagnetic (EM) field itself and quantized  field of lattice deformations (phonon field) will be in the nearest future the working components of optoelectronic,  spintronic devices and various logic quantum systems including quantum computers and quantum communication systems.

We will also  show in present work  the other ways to obtain long-lived coherent states with similar or even more extensive field of practical application to be consequence of  more long lifetimes. Main difference of given states, that they are collective coherent states in comparison with individual coherent states, realised on point centers, which are subject of many theoretical and experimental studies, known from literature.

The aim of the work presented is to review briefly the theoretical and experimental results, based mainly on the works of authors, in the application of QFT to elaboration of new ways of long-lived coherent states' formation and its experimental confirmation on the example of  carbon low-dimensional systems - quasi-1D carbon nanotubes, carbynes and graphene.

The simplest models which capture the
salient features of the relevant physics in   field  of practical application  of QED are
the Jaynes-Cummings model (JCM) \cite{Jaynes_Cummings} for the one qubit
case and its generalization for multiqubit systems by
Tavis and Cummings \cite{Tavis}. Tavis-Cummings model was generalized in \cite{Slepyan_Yerchak} (SYHB-model), by taking into account the 1D-coupling between qubits. Recently QED-model for  single chain coupled qubit system was generalized 
  for quasionedimensional axially symmetric  multichain coupled qubit system  \cite{Dovlatova_A_Yearchuck_D}. 
  It is substantial, that in the model, proposed in \cite{Dovlatova_A_Yearchuck_D}  (DY-model) the interaction of quantized EM-field with multichain qubit system  is considered by taking into account both the intrachain and interchain qubit coupling without restriction on their values.   From theoretical results in \cite{Slepyan_Yerchak}, \cite{Dovlatova_A_Yearchuck_D}, \cite{Dovlatova_Dmitri} and from their experimental confirmation in 
 \cite{Yerchuck_D_Dovlatova_A} follows, that by strong interaction of EM-field with matter the correct description of spectroscopic transitions including stationary spectroscopy is achieved the only in the frame of QED consideration. It  concerns both optical and radio spectroscopies, that means, that QED consideration has to be also undertaken by
 electron spin resonance (ESR) studies in the case of strong interaction of EM-field with spin systems. The review of given results is given in Section 2. 

 It is reasonable to suggest, that  analogous conclusion can be
drawn for the case of strong interaction of phonons with spin system or electron system. In other words it seems to be reasonable the idea, that relaxation of paramagnetic (or optical) centers in the case of strong spin-phonon (electron-phonon) interaction can be described correctly the only in the frames of quantum field theory.  The system of equations for  dynamics of spectroscopic transitions in 1D multiqubit  exchange coupled  (para)magnetic and optical systems by strong dipole-photon and dipole-phonon coupling  within the framework of quantum electrodynamics and quantum deformation field theory {phonon  theory} was  derived in \cite{Dovlatova_Yerchuck}. It was  showed, that new quantum physics phenomenon - the formation of longlived coherent state of joint systems \{electric(magnetic) dipoles + resonance phonons\}, leading  to appearance of quantum acoustic (phonon) Rabi oscillations has to be taking place. The results obtained are reviewed in Section 3.

Let us also remark, that in all theoretical and experimental studies of the matter systems by means of stationary optical- and radio spectroscopy methods EM-field is considered the only classically. For example, in  relatively recent work \cite{D_W_Wang} the review of development in the theory of resonant  Raman
scattering (RS) in 1D electron systems is given by using several different quantum theoretical models of the matter subsystem - Fermi liquid model, Luttinger liquid model
and Hubbard model, however  by classical description of EM-field.

 \section{Longlived Coherent State Formation in Carbon Low-Dimensional Materials in  Result of Rabi-Wave Packets' Propagation}

 QED-model for  multichain coupled qubit system, that is DY-model,  proposed in \cite{Dovlatova_A_Yearchuck_D} predicts, that by strong electron-photon interaction quantum nature of EM-field can become apparent in any stationary optical experiments. In particular, new quantum optics phenomenon - Rabi waves and Rabi wave packets' formation, which was theoretically predicted for the fist time in \cite{Slepyan_Yerchak}, can give essential contribution in stationary spectral distribution of Raman
scattering (RS) intensity and  spectral distributions of infrared (IR), visible or ultraviolet absorption, reflection, transmission intensities.
The calculation \cite{Dovlatova_A_Yearchuck_D} is illustrated on the example of perfect
quasi-one-dimensional carbon zigzag shaped carbon nanotubes (CZSNTs).
In fact 2D-1D transition appearance  (new quantum size  effect) in physical properties of carbon nanotubes, which is realised with diameter
decrease, is theoretically argued. It is predicted, that 2D-1D
transition leads to necessity of qualitatively different electronic model for quasi-1D CZSNTs with
 cardinal change of their physical properties. The electronic model for quasi-1D CZSNIt is the following.
 The single quasi-1D CZSNT represents itself autonomous dynamical system with discrete circular symmetry, consisting of finite number $n \in N$ of carbon backbones of trans-polyacetylene (t-PA) chains, which are placed periodically along transverse angle coordinate. Longitudinal axes $\{x_i\}$, $i = \overline(1,n)$, of individual chains can be directed both along element of cylinder and along generatrix of any other smooth figure with axial symmetry. It is similar to Su, Schrieffer, Heeger (SSH) \cite{Su_Schrieffer_Heeger_1979}, \cite{Su_Schrieffer_Heeger_1980} model of 1D organic conductors in the part, concerning the choose of active
degrees of freedom, that allows to consider all $n$ carbon chains in the model of quasi-1D CZSNTs to be equivalent each other, while in real quasi-1D CZSNTs the adjacent
chains represent themselves a mirror of each other relatively corresponding planes, passing through NT-axis.
Given $n$-chain  set can be  considered to be a single whole, which  holds the quasi-one-dimensionality of a single chain. It seems to be correct for perfect CZSNTs, if their diameter is $\leq$ 1 nm . 

At the same time it is well known, that free standing nanotubes are considered theoretically   to be $2D$-strutures and two-dimensional lattice structure of a single wall carbon nanotube (SWNT)
is determined by the chirality, which is defined by two integers
$(n,m)$ \cite{Dresselhaus}, \cite{Saito}.  Two in-plane $G$ point longitudinal and transverse optical phonon $(LO$ and $TO)$ modes \cite{Reich} and
the out-of-plane radial breathing mode $(RBM)$ \cite{Dresselhaus M.S} are observed in the Raman spectra of  SWNTs.
The $LO$ and $TO$ phonon modes at
the $G$ point in the two-dimensional Brillouin zone
are degenerate in 3D-graphite and 2D-graphene, however they split in SWNTs into two peaks,
denoted by $G^+$ and $G^-$peaks, respectively, \cite{G.Dresselhaus}. It is  consequence of the
curvature effect. The agreement of experimental Raman studies of carbon nanotubes (NTs) with diameter $ \geq 1 nm$ with $2D$ SWNT-theory unambiguously indicates, that given NTs, produced mainly by various CVD-methods  are really $2D$ systems (in cylindric space).  At the same time the narrow  NTs with diameter $ < 1 nm$ cannot be considered strongly speaking to be  $2D$-systems, they are quasi-$1D$ systems, and in \cite{Dovlatova_A_Yearchuck_D} and in  \cite{Yerchuck_D_Dovlatova_A} is argued,  that 2D-1D transition  (quantum  size effect) has to be take place and  it really  takes place by its control with radiospectroscopy methods.

At the same time the presence of both    intrachain and interchain qubit coupling by simultaneous preservation of quasionedimensionality seems to be leading to additional requirements for observation of size effects by 2D-1D transition, especially  with optical methods, which are very sensitive to structural perfectness. It is understandable, that  along with requirement of quasionedimensionality the  additional requirement of homogeneity of the ensemble of NTs arises.  It means, that any dispersion in axis direction, chirality, length and especially in diameter both for single NT along its axis and between different NTs in ensemble has to be absent, axial symmetry has to be also retained, that is, there are additional requirements in comparison with, for example, t-PA technology. The CVD-technology of NTs production and many similar to its seem to be not satisfying to above-listed requirements at present. It means, that experimental results and their theoretical treatment will be different in both the cases, that really takes place. The situation seems to be analogous to some extent to the solid state physics of the same substance in single crystal- and amorphous forms. It was shown in \cite{Yerchuck_D_Dovlatova_A}, that the technology of CZSNTs formation, based on  high energy ion beam modification (HEIBM) of natural diamond single crystals,  satisfy given requirements. 
 
Semiclassical evaluation, obtained in \cite{Dovlatova_A_Yearchuck_D}, has led to  conclusion, that $\pi$-subsystem of quasi-1D CZSNTs will be inactive in optical experiments. The comparison with optical properties of related carbon chain material - carbynoids - has allowed to predict, that among possible optically active topological defects in  $\sigma$-subsystem of quasi-1D CZSNTs the $\sigma$-polaron is expected to be prevailing. 

Each $\sigma$-polaron interacting with external EM-field in accordance with experiment in \cite{Yearchuck_PL} can be approximated like to guantum dots in \cite{Slepyan_Yerchak} by two-level qubit. Then  the Hamiltonian, proposed in the work \cite{Slepyan_Yerchak} can be   generalized, that was done.  The insufficient for the model local field term was omitted. (Local field term seems to be playing minor role by description of $\sigma$-polarons in comparison with quantum dots, since size of quantum dots is greatly exceeding the size of $\sigma$-polarons).  The apparatus of hypercomplex $n$-numbers was used.  Hypercomplex $n$-numbers are defined to be elements of commutative ring, given by (\ref{Eq18})  
\begin{equation}
\label{Eq18}
Z_n = C \oplus {C} \oplus{...}\oplus {C},
\end{equation}
$Z_1 = C$.
that is, it is direct sum of $n$ fields of complex numbers $C$, $n\in N$. It means, that any hypercomplex $n$-number $z$ is $n$-dimensional quantity with the components $k_\alpha \in C$, $\alpha = \overline{0,n-1}$, that is in row matrix form $z$ is
\begin{equation}
\label{Eq19}
z = [k_0, k_1, k_2, ..., k_{n-1}].
\end{equation}
Hypercomplex $n$-number $z$ can be represented also in the form
\begin{equation}\label{Eq20}
z = \sum_{\alpha = 0}^{n-1}k_\alpha\pi_\alpha,
\end{equation}
where $\pi_\alpha$ are basis elements of $Z_n$. They are
\begin{equation}
\label{Eq21}
\begin{split}
&\pi_0 = [1,0, ...,0,0],  \pi_1 =[0,1, ...,0,0],\\
&..., \pi_{n-1} = [0,0, ...,0,1].
\end{split}
\end{equation}
In other words, the set of $k_\alpha \in C, \alpha = \overline{0, n-1}$ is the set of eigenvalues of hypercomplex $n$-number $z \in Z_n$, the set  of $\{\pi_\alpha\}$, $\alpha = \overline{0, n-1}$ is eigenbasis of $Z_n$-algebra.
Then the QED-Hamiltonian, considered to be hypercomplex operator $n$-number, for $\sigma$-polaron system of interacting with EM-field CZSNTs, consisting of $n$ backbones of $t$-PA chains, which are connected between themselves in that way, in order to produce rolled up graphene sheet, in matrix representation is \begin{equation} 
\label{Eq10a}
[\hat{\mathcal{H}}] = [\hat{\mathcal{H}}_{\sigma}] + [\hat{\mathcal{H}}_F] + [\hat{\mathcal{H}}_{\sigma F}] + [\hat{\mathcal{H}}_{\sigma \sigma}].
\end{equation}
The rotating wave approximation and  the single-mode approximation of EM-field are used. All the components in (\ref{Eq10a}) are considered to be hypercomplex operator $n$-numbers and they are the following. $[\hat{\mathcal{H}}_{\sigma}]$ represents the operator of the  energy of $\sigma$-polaron subsystem  in the absence of interaction between $\sigma$-polarons  themselves and with EM-field. It is
\begin{equation} 
\label{Eq11a}
[\hat{\mathcal{H}}_{\sigma}] = (\hbar \omega _0/2) \sum_{j = 0}^{n-1}\sum_m {\hat {\sigma}^z_{mj}}[e_1]^j,
\end{equation}
  where $\hat {\sigma}^z_{mj} = \left|a_{mj}\right\rangle  \left\langle a_{mj} \right|-\left|b_{mj}\right\rangle  \left\langle b_{mj} \right|$ is $z$-transition operator between the ground and excited states of $m$-th $\sigma$-polaron in $j$-th chain. The second term 
\begin{equation} 
\label{Eq12a}
[\hat {\mathcal{H}}_F] = \hbar \omega \sum_{j = 0}^{n-1}\hat {a}^+\hat {a}[e_1]^j 
\end{equation}
is the Hamiltonian of the free EM-field, represented in the form of 
hypercomplex operator $n$-number.
The component of the Hamiltonian (\ref{Eq10a})
\begin{equation}
\label{Eq13a}
[\hat {\mathcal{H}}_{\sigma F}] =\hbar g \sum_{j = 0}^{n-1}\sum\limits_m {(\hat {\sigma }_{mj}^+\hat {a}e^{ikma} + \hat {\sigma }_{mj}^-\hat {a}^+e^{-ikma})}[e_1]^j 
\end{equation}
corresponds  to the interaction of $\sigma$-polaron sybsystem with EM-field, where $g$  is the interaction constant.
The next term in Hamiltonian (\ref{Eq10a})
 describes the intrachain and interchain $\sigma$-polaron–$\sigma$-polaron  interaction. It is given by the
expression
\begin{equation}
\begin{split}
\label{Eq14a}
&[\hat{\mathcal{H}}_{\sigma\sigma}] =
-\hbar \sum_{l = 0}^{n-1}\sum_{j = 0}^{n-1}\xi^{(1)}_{|l-j|}[e_1]^l\sum\limits_m \left|a_{mj} \right\rangle \left\langle a_{m+1,j}\right|[e_1]^j \\
&-\hbar\sum_{l = 0}^{n-1}\sum_{j = 0}^{n-1}\xi^{(1)}_{|l-j|}[e_1]^l\sum\limits_m \left|a_{mj} \right\rangle \left\langle a_{m-1,j}  \right| [e_1]^j\\
&-\hbar\sum_{l=0}^{n-1}\sum_{j = 0}^{n-1}\xi^{(2)}_{|l-j|}[e_1]^l\sum\limits_m  \left| b_{mj} \right\rangle \left\langle b_{m+1,j}\right|[e_1]^j\\ &-\hbar\sum_{l=0}^{n-1} \sum_{j = 0}^{n-1}\xi^{(2)}_{|l-j|}[e_1]^l\sum\limits_m\left| b_{mj} \right\rangle \left\langle b_{p-1,j} \right|[e_1]^j,
\end{split}
\end{equation}
where  $\hbar\xi^{(1,2)}_{|l-j|}$ are the energies, characterising intrachain $(l = j)$ and interchain $(l \neq j)$  $\sigma$-polaron–$\sigma$-polaron interaction for the excited ($\xi^{(1)}$) and ground ($\xi^{(2)}$) states of $j$-th chain, $\left| b_{mj} \right\rangle, \left|a_{mj}\right\rangle$ are ground and excited states correspondingly of $m$-th $\sigma$-polaron of $j$-th chain. Matrix $[e_1]^j$ in (\ref{Eq11a}) to (\ref{Eq14a})  is j-th power of the circulant matrix $[e_1]$, which is the following
\begin{equation}
\label{Eq16a}
[e_1]=\left[\begin{array} {*{20}c} 0&1&0& ...&0  \\ 0&0&1& ...&0 \\ &...& \\ 0&0& ... &0&1\\1&0&...&0&0 \end{array}\right].
\end{equation}

 The solution of nonstationary Schr\"odinger equation for hypercomplex matrix function, determined by 
\begin{equation}
\begin{split}
\label{Eq17a}
&[\left| {\Psi (t)} \right\rangle] = \\ 
&\sum_{j = 0}^{n-1}\{\sum\limits_l \sum\limits_{m} \left(A^j_{m,l}(t) \left|a_{mj},l \right\rangle + B^j_{m,l}(t) \left| b_{mj},l \right\rangle\right) \}[e_1]^j.
\end{split}
\end{equation}
where $\left| b_{mj},l \right\rangle = \left| b_{mj} \right\rangle\otimes\left|l \right\rangle$, $\left| a_{mj},l \right\rangle = \left| a_{mj} \right\rangle\otimes\left|l \right\rangle$,   $\left|l \right\rangle$ is the EM-field  Fock state with $l$  photons, $A^j_{m,l}(t)$, $B^j_{m,l}(t)$ are the unknown probability amplitudes,
  was obtained   in continuum limit $([\left| {\Psi (t)}_{cont} \right\rangle] \equiv  [\left|\Phi^l(x,t)\right\rangle])$ and it was represented in the form of sum of $n$ solutions for $n$ chains, that is, hypercomplex $n$-number $\Phi^l(x,t) $ is
\begin{equation}
\label{Eq30a}
\Phi^l(x,t) = \sum_{q = 0}^{n-1}\tilde{\Phi}^l_q(x,t),
\end{equation}
where the solution for $q$-th chain $\tilde{\Phi}^l_q(x,t)$ is
\begin{equation}
\label{Eq31a}
\tilde{\Phi}^l_q(x,t) = \sum_{p = 0}^{n-1}\Phi^l_{qp}(x,t)[e_1]^p,
\end{equation}
in which the  matrix elements $\Phi^l_{qp}(x,t)$   of matrix  $[\Phi^l(x,t)]$ are
\begin{equation}
\begin{split}
\label{Eq26a}
&\Phi^l_{qp}(x,t) = \int\limits_{-\infty }^{\infty}\Theta^l_{q}(h,0) \exp{ \frac{-2\pi qpi}{n}} \exp{ihx}\times \\
&\exp{\{i\sum_{j = 0}^{n-1}\exp{\frac{2\pi qji}{n}(\vartheta_j(h) - g\sqrt{l-1}\kappa_j(h))}\}}dh,
\end{split}
\end{equation}
where $\Theta^l_{q}(h,0)$, $\vartheta_j(h)$, $\kappa_j(h)$ are determined by eigenvalues $\textbf{k}_\alpha \in C, \alpha = \overline{0, n-1}$ of $\Phi^l(h,0)$, $\theta(h)$ and $\chi(h)$, which are considered to be hypercomplex $n$-numbers. They are
\begin{equation}
\label{Eq27a}
\Theta^l_{q}(h,0) = \frac{1}{n}\textbf{k}_q (\Phi^l(h,0)) = \frac{1}{n}\sum_{j = 0}^{n-1}\Phi_j^l(h,0)\exp{\frac{2\pi q j i}{n}}
\end{equation}
\begin{equation}
\label{Eq28a}
\vartheta_j(h) = \frac{1}{n}\textbf{k}_j(\theta(h)) = \frac{1}{n}\sum_{r = 0}^{n-1}\theta_r(h)\exp{\frac{2\pi  jr i}{n}},
\end{equation}
\begin{equation}
\label{Eq29a}
\kappa_j(h) = \frac{1}{n}\textbf{k}_j(\chi(h)) = \frac{1}{n}\sum_{r = 0}^{n-1}\chi_j(h)\exp{\frac{2\pi j r i}{n}}.
\end{equation}
It was taken into account, that for the state vector $[\left| {\Psi (t)} \right\rangle]_cont$   in continuum limit we have
\begin{equation}
\begin{split}
\label{Eq18a}
&[\Phi^l(x,t)] = \\
&\int\limits_{-\infty }^\infty{[\overline{\Phi}^l(h,0)]\exp\{i t([\theta^l(h)] - g \sqrt{l+1}[\chi])\}e^{ihx}}dh,
\end{split}
\end{equation}
where $x$ is  hypercomplex axis $x = [x, x, ..., x]$, $[\Phi^l(x,t)]$ is 
\begin{equation}
\begin{split}
\label{Eq19a}
[\Phi^l(x,t)] = exp{\frac{i(\omega_0 t - kx)[\sigma_z]}{2}}\exp{\frac{\lambda t}{2}}[\Psi^l(x,t)], 
\end{split}
\end{equation}
 In its turn $[\Psi^l(x,t)]$ is continuous limit of functional block matrix of discrete variable $m$, which is given by (\ref{Eq17ef}).
\begin{equation}
\begin{split}
\label{Eq17ef}
[\Psi_{m,l}(t)] = \left[\begin{array} {*{20}c}&[A_{m,l}(t)]\\
&      \\
&[B_{m,l+1}(t)]\end{array}\right],  
\end{split}
\end{equation} 
consisting of two $[n \times n]$ matrices of probability amplitudes
\begin{equation}
\begin{split}
\label{Eq21a}
&[A_{m,l}(t)] = \sum_{j = 0}^{n-1}A^j_{m,l}(t)[e_1]^j, \\
&[B_{m,l+1}(t)] = \sum_{j = 0}^{n-1}B^j_{m,l+1}(t)[e_1]^j,
\end{split}
\end{equation}
which are determined by relationship (\ref{Eq17a}).
Further, matrix $[\theta(h)]$ in (\ref{Eq18a}) is 
\begin{equation}
\begin{split}
\label{Eq22a}
&[\theta(h)] = \frac{1}{2}\{([\theta_1(h)] + [\theta_2(h)]) \otimes [E_2]\} \\
&+ \frac{1}{2}\{([\theta_1(h)] - [\theta_2(h)]) \otimes [\sigma_z]\}, 
\end{split}
\end{equation}
where  $[\theta_1(h)]$ and $[\theta_2(h)]$ are 
\begin{equation}
\begin{split}
\label{Eq23a}
&[\theta_1(h)] = [\xi_1]\{2 - a^2(h + \frac{k}{2})^2\}, \\
&[\theta_2(h)] = [\xi_2]\{2 - a^2(h - \frac{k}{2})^2\}
\end{split}
\end{equation}
Matrix $[\chi]$ in (\ref{Eq18a}) is
\begin{equation}
\begin{split}
\label{Eq25a}
[\chi] = [E_n] \otimes [\sigma_x]\exp({-i [\sigma_z] (\omega - \omega_0) t}),
\end{split}
\end{equation}
where $[E_n]$ is $[n \times n]$ unit matrix.

The relationship (\ref{Eq31a}) by taking into account  (\ref{Eq26a}) - (\ref{Eq25a}) determines Rabi-wave packet, which propagates along individual chain of  CZSNT.  Subsequent analysis of Rabi-wave packet dynamics for individual chain in CZSNT will be now coinciding by rescaling of parameters with analysis of Rabi-wave packet dynamics in quantum dot chain, considered in \cite{Slepyan_Yerchak}. In particular, in the same manner can be obtained temporal dependence of the integral inversion, which gives in implicit form  the way for  comparison of theoretical results with any stationary optical experiments in any quasi-1D system with strong electron-photon interaction. It is sufficient to make a Fourier transform of  given temporal dependence. Fourier transformed temporal dependence of the integral inversion will be proportional to spectral distribution by IR-absorption, IR-transmission, IR-reflection or Raman scattering, since they are  determined by population difference. It means, that dynamical quantum nonstationary properties of optical systems can become apparent by conventional stationary registration of the spectra. 

Given conclusion was confirmed in \cite{Yerchuck_D_Dovlatova_A} by the RS study of quasi-1D CZSNTs, IR study of carbynes and by analysis of RS results in graphene.

Samples of type IIa natural diamond, implanted by high energy ions of copper and boron (the energy of ions in ion beam is $63$ $MeV$ and $13.6$ $MeV$ for copper and boron ions correspondingly, ion beam dose is $5\times{10^{14}}$ $cm^{-2}$)  have been studied in given work. Both the samples represent in their geometry prismes with quilateral triangle in their bases, coinciding with (111) crystallographic plane,  side of base triangle was equal to $\approx 5 mm$, depth of the samples was equal to $\approx 1 mm$. Ion implantation was performed along $\left\langle{111}\right\rangle$ crystal direction, that is transversely to triangle prism base uniformly along all the surface. 

Raman scattering  (RS) spectra were registered in backscattering geometry. Laser excitation wave length was 488 $nm$, rectangular slit $350{\times}350 (\mu m)^2$ was used, scan velocity was 100 $cm^{-1}$ pro minute.

Infrared  absorption  and reflection studies have been performed on uniaxially oriented
 carbynoid film samples
prepared by chemical dehydrohalogenation of poly(vinylidene fluoride) (PVDF), which were stored 12 Y at room temperature.
Preparation details of carbynoids were described in \cite{Ertchak_J_Physics_Condensed_Matter}. The FTIR spectrometer "Nexus" has been used, IR  spectra were registered in the range $400 - 5000$ $cm^{-1}$ at room temperature. Two groups of uniaxially oriented
 carbynoid film samples were studied, designated correspondingly $A$  and $B$ samples. Carbynoid samples con-
tained a rather high concentration of residual fluorine
and technological oxygen atoms. The samples of $A$ set
(hereinafter $A$-samples, their  designation in \cite{Ertchak_J_Physics_Condensed_Matter} is the sec-
ond series samples) were with $F/C$ ratio equal to
3/7, their oxygen contamination $O/C$ was 1 to 5. $A$-samples were thermally treated at 120°C for 2 hours.
The contamination of fluorine and oxygen atoms in the
samples of $B$ series (hereinafter $B$-samples, their  designation in \cite{Ertchak_J_Physics_Condensed_Matter} is the  third series samples) was intermediate
between 3/10 and 3/7 for the $F/C$ ratio and between 1
to 10 and 1 to 5 for $O/C$ ratio (however O and F content
was not determined exactly). Like to the classification
proposed for doped t-PA \cite{Heeger_1988}, the samples studied can
be attributed both to highly doped carbynes and to carbynoids,
that is to materials including a wide range of carbyne-like
structures. It was established, that the difference between the spectra for the samples, belonging to the same group is in limits of uncertainty of experimental measurements, and the spectra for two samples belonging to different groups was presented and analysed in \cite{Yerchuck_D_Dovlatova_A}.

\begin{figure}
\includegraphics[width=0.5\textwidth]{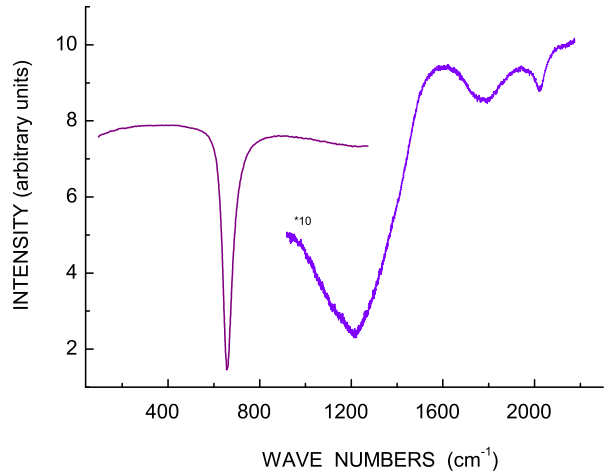}
\caption[Spectral distribution of Raman scattering  intensity in
diamond single crystal, implanted by high energy copper ions, the excitation is from implanted side of the sample]
{\label{Figure1} Spectral distribution of Raman scattering  intensity in
diamond single crystal, implanted by high energy copper ions, the excitation is from implanted side of the sample}
\end{figure}

 \begin{figure}
\includegraphics[width=0.5\textwidth]{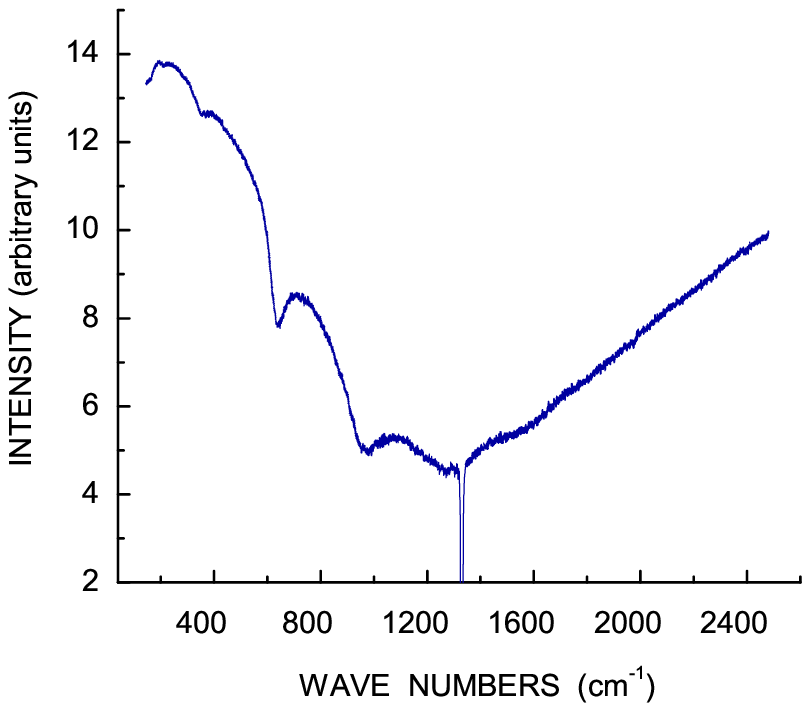}
\caption[Spectral distribution of Raman scattering  intensity in
diamond single crystal, implanted by high energy copper ions, the excitation is from unimplanted side of the sample]
{\label{Figure2} Spectral distribution of Raman scattering  intensity in
diamond single crystal, implanted by high energy copper ions, the excitation is from unimplanted side of the sample}
\end{figure}

\begin{figure}
\includegraphics[width=0.5\textwidth]{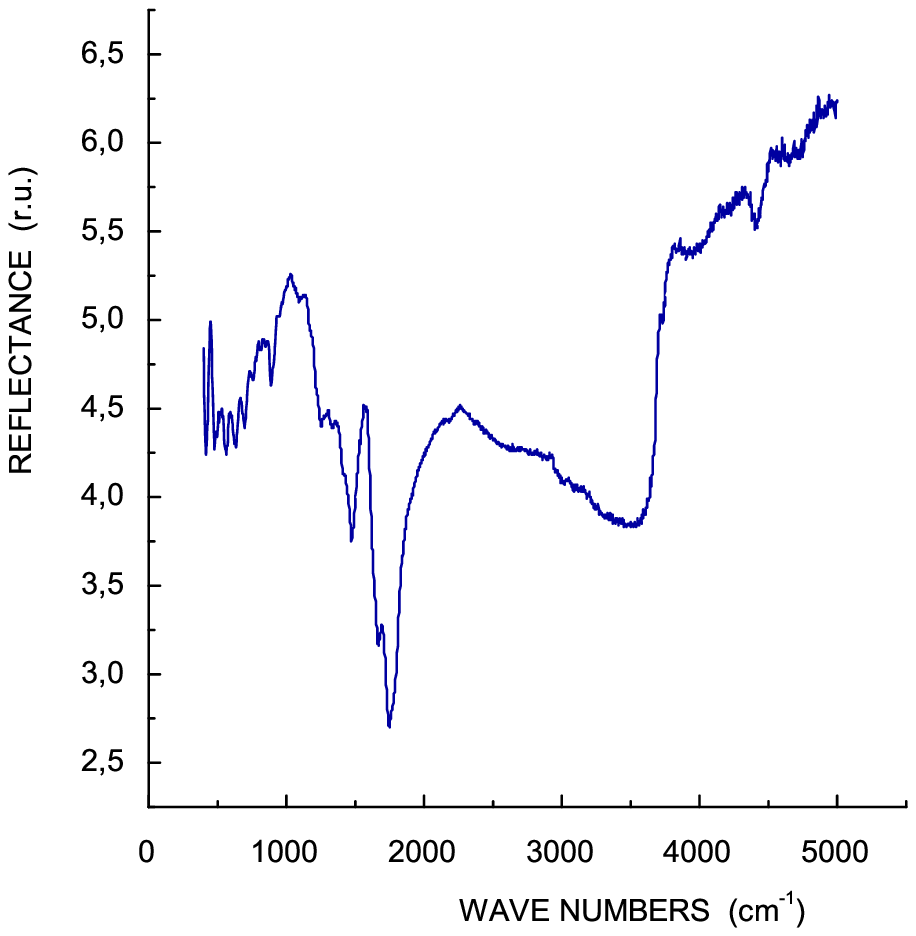}
\caption[Spectral distribution of IR reflection intensity in carbynoid uniaxially oriented $A$-sample]
{\label{Figure3} Spectral distribution of IR reflection intensity in carbyne uniaxially oriented $A$-sample}
\end{figure}
\begin{figure}
\includegraphics[width=0.5\textwidth]{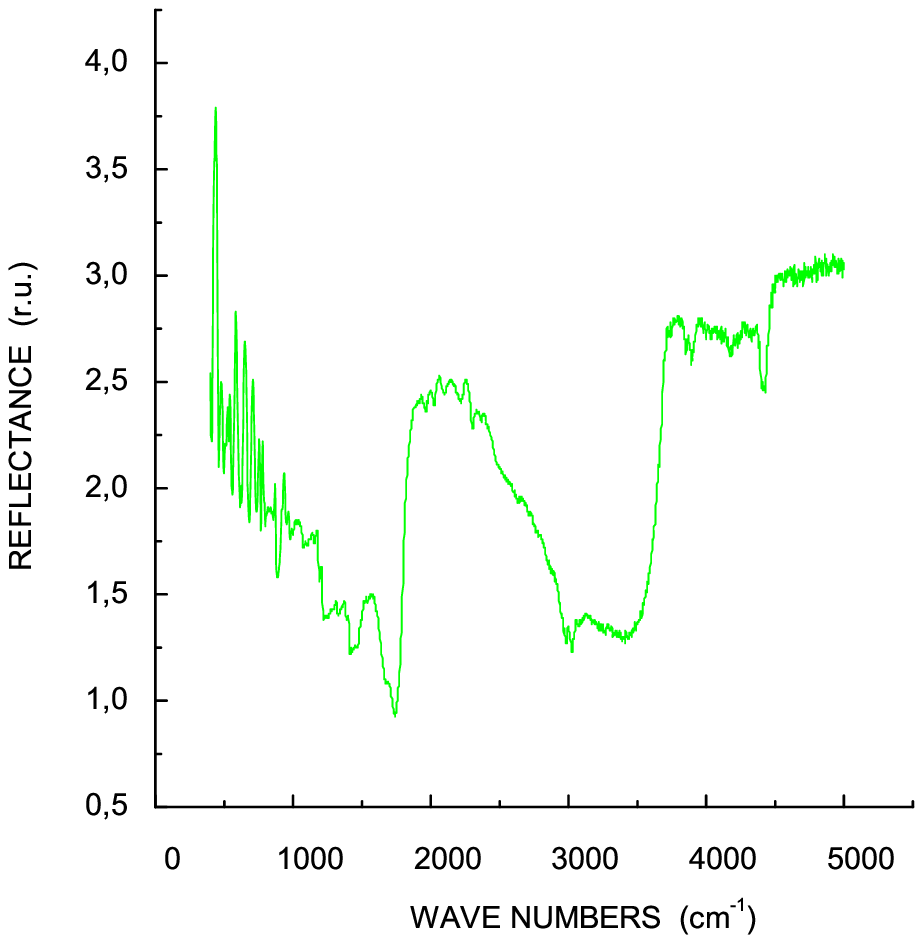}
\caption[Spectral distribution of IR reflection intensity in carbynoid uniaxially oriented $B$-sample]
{\label{Figure4} Spectral distribution of IR reflection intensity in carbyne uniaxially oriented $B$-sample}
\end{figure}
\begin{figure}
\includegraphics[width=0.5\textwidth]{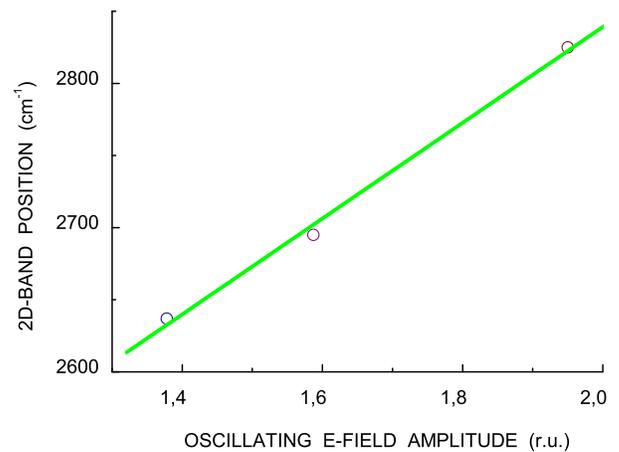}
\caption[Dependence of $2D$-band position in graphene on electrical component of oscillating excitation field]
{\label{Figure5} Dependence of $2D$-band position in graphene on electrical component of oscillating excitation field}
\end{figure}

 It is seen from Figures 1,  that Raman spectra  are characterised in quasi-1D CZSNTs, produced by $\langle{111}\rangle$ HEIBM of diamond single crystals (in copper implanted sample by excitation from impanted side of the sample) by  the only  single vibronic mode of the lattice, formed by Su-Schrieffer-Heeger $\sigma$-polarons with peak position at $656.8{\pm}0.2$ $cm^{-1}$ and by additional lines with peak positions $1215{\pm}1$ $cm^{-1}$, $1779.5{\pm}1$ $cm^{-1}$,   $2022.3{\pm}0.5$ $cm^{-1}$ corresponding to Fourier transform of revival part  of of the time-dependence of integral inversion, detection of which in stationary RS-measurements is determined by formation and propagation of quantum Rabi  wave  packet, to be consequence of quantum nature of EM-field. Given RS-mode is appeared instead of longitudinal
and transverse optical phonon    $G^+$ and $G^-$modes and
the out-of-plane radial breathing mode, which are observed in Raman spectra of  2D single wall nanotubes \cite{G.Dresselhaus}.

The identification of the lines $1215 {\pm} 1$ cm$^{-1}$, $1779.5 {\pm} 1$ cm$^{-1}$,  $2022.3{\pm}0.5$ cm$^{-1}$ (and corresponding lines in boron ion modified sample) with Fourier-image of revival part of Rabi packet is confirmed by the following. It is well known \cite{Abragam}, that  in the case of point absorbing centers classical Rabi frequency is linear function of the amplitude of oscillating EM-field. It was experimentally confirmed in \cite{Yerchak}, \cite{Yerchak_Stelmakh}. Given dependency takes also place for the center of Rabi wave packet, that follows from the analysis of Fourier transform of temporal dependence  of the  integral inversion. Further, it is evident, that amplitude of electrical component of laser irradiation, penetrating in ion beam modified region (IBMR), is lesser by excitation from unimplanted side. It is consequence of some absorption in the unimplanted volume of diamond crystal. We see, that really, the frequency values of additional two rightmost  Raman peaks at 1569 ${\pm}3$ cm$^{-1}$, 1757 ${\pm} 5$ cm$^{-1}$, which are observed in Raman spectrum  by excitation of  diamond sample with incorporated CZSNTs from unimplanted side are substantially less, than the frequency values of additional   two rightmost spectral components at $1779.5 {\pm} 1$ cm$^{-1}$ and  $2022.3{ \pm} 0.5$ cm$^{-1}$, observed by excitation from implanted side, compare Figure 2 and  Figure 1. Relative frequency changes are 1,151 and 1.134 ${\pm}0.003$ for the pairs [$2022.3 {\pm} 0.5$ cm$^{-1}$, 1757 ${\pm} 5$ cm$^{-1}$] and [$1779.5 {\pm} 1$ cm$^{-1}$, 1569 ${\pm} 3$ cm$^{-1}$] correspondingly and it is seen, that they are close to each other. It is substantial,  that more high frequency undergoes  slightly more large relative change in correspondence with theoretical analysis.  

 The lines at 354.6,  977.1 (${\pm}1$) cm$^{-1}$, Figure 2, were assigned with two antiferroelectric spin wave resonance (AFESWR) modes. It means, taking into account the conclusion on some magnetic ordering quasi-1D CZSNTs in \cite{Erchak} and direct  observation in the same sample of ferromagnetic spin wave resonance (FMSWR) \cite{Ertchak}, \cite{Ertchak_Stelmakh}, that quasi-1D CZSNTs are multiferroic materials. 
Let us remark, that 
AFESWR is new optical phenomenon, which was theoretically described and experinmentally confirmed for the first time in \cite{Yearchuck_PL} by the study of optical properties of carbynes. It seems to be very interesting, that  carbynes along with antiferroelectric ordering possess also by  ferroelectric and ferromagnetic ordering. The conclusion on ferromagnetic ordering has been obtained  from ESR studies  of  heavily doped by technological impurities carbynes in \cite{Ertchak_J_Physics_Condensed_Matter}, \cite{Ertchak_D} by means of immediate observation of ferromagnetic spin wave resonance.  The  optical studies  of the same samples in \cite{Kudryavtsev_Yearchuck}, \cite{Yearchuck_Strasbourg}, \cite{Yearchuck_Wiesbaden}, \cite{Yearchuck_ArXiv}, \cite{Yearchuck_PL}, \cite{Yearchuck_Yerchak_Doklady} allowed to estabish the origin and the structure of infrared (IR) and Raman scattering (RS) active centers, at that the conclusion on both antiferroelectric ordering  and simultaneous ferroelectric ordering of  carbynes was obtained directly by registration of  antiferroelectric spin wave resonance and ferroelectric spin wave resonance (FESWR) in \cite{Yearchuck_ArXiv}, \cite{Yearchuck_PL} and \cite{Yearchuck_Yerchak_Doklady} respectively. Especially interesting, that both antiferroelectricity,  ferroelectricity and ferromagnetism have been observed simultaneously in the same samples and, what is characteristic, in the same temperature range, in particular, at room temperature. Let us remark, that along with antiferroelectric spin wave resonance ferroelectric spin wave resonance is also new quantum physics phenomenon and it has been identified in \cite{Yearchuck_Yerchak_Doklady} for the first time in optical spectroscopy too.  Let us also remark,  that given phenomenon and its characteristic features were predicted theoretically in \cite{Yerchak_Yearchuck}. 

Antiferroelectricity  seems to be very new property of pure carbon at all.  Moreover antiferroelectric  ordering has been found in the matter for the first time (to our knowlegde). 
 
Let us remark, that experimental observation of multiferroicity in quasi-1D CZSNTs means the breakdown of space inversion symmetry  along hypercomplex CZSNT symmetry axis $x$. It agrees well with the model of quasi-1D CZSNTs, \cite{Dovlatova_A_Yearchuck_D},   \cite{Yerchuck_D_Dovlatova_A}, based on bond dimerisation in  all chain components of  quasi-1D CZSNT along its symmetry axis $x$, which leads to inversion symmetry breakdown along given axis. Therefore, the experimental observation of antiferroelecricity of quasi-1D CZSNTs, necessary condition of which is the evident prediction of the model, proposed in \cite{Dovlatova_A_Yearchuck_D}, can be considered to be additional argument in favour of given model.
  
At the same time, typical multiferroics belong to the group of the perovskite transition metal 
oxides, and include rare-earth manganites and -ferrites, that is, they are the compounds, which include atoms with unfilled inner $d$ or $f$ atomic shells (for example $TbMnO_3$, $HoMn_2O_5$, 
$LuFe_2O_4$). Other examples are the bismuth compounds $BiFeO_3$ and $BiMnO_3$, and non-oxides 
like to $BaNiF_4$ and spinel chalcogenides, for example $ZnCr_2Se_4$. These alloys show rich 
phase diagrams combining different ferroic orders in separate phases. Apart from 
single phase multiferroics, composites and heterostructures exhibiting more than one ferroic order parameter are existing. Given comparison indicated, that mechanism of multiferroicity is quite other, which is the  subject 
of another original research.

The substantial decrease of relative intensity of 641.8 ${\pm}1$ $cm^{-1}$ mode in comparison with  $656.8{\pm}0.2$ cm$^{-1}$ mode testifies in favour of given assignment. It is seen, that AFESWR-splitting is rather large and it has the same order of values with the splitting between two $\sigma$-polaron vibronic levels. It means, that linear AFESWR-theory  \cite{Yearchuck_PL},  which predicts   a set of equidistant  AFESWR-modes, arranged the left and the right of central mode, can be used the only to obtain approximately the average value of AFESWR-splitting. Really, it is seen from Figure 2, that in given case AFESWR-modes   are not equidistant, they are shifted on different distances 335.3, 287,2 cm$^{-1}$ from main AFESWR-mode. At the same time average value of AFESWR-splitting is 311.3 cm$^{-1}$ and it is close to the value in 300 cm$^{-1}$  expected in accordance with linear AFESWR-theory by taking into account experimental AFESWR-splitting value in carbynes \cite{Yearchuck_PL}. 

  Therefore, in \cite{Yerchuck_D_Dovlatova_A}, the direct proof of assignment of the lines 656.8 ${\pm} 0.2$ cm$^{-1}$ and 641.8 ${\pm} 1$ cm$^{-1}$ with localised vibration mode of $\sigma$-polaron lattice was obtained.  
 
 To explain  differences in RS spectral characteristics by the change of the direction of the excitation wave propagation, we have to take into consideration the following. The possibility of spin wave resonance excitation in 1D-systems is strongly dependent on  the geometry of experiment, determined by  directions of chain axis, vectors of  external static  magnetic field (intracrystalline electric field)  and oscillating magnetic (electric) fields by FMSWR (FESWR) study correspondingly. It is confirmed by FMSWR study in carbynes, \cite{Ertchak_J_Physics_Condensed_Matter}. The shape of CZSNTs is not strictly cylidric  in the end of ion run and the axes $x_i$, $i = \overline{1,n}$ are generatrixes  of the figure of onion-like shape, that provides the necessary geometry for  AFESWR-excitation on $\sigma$-polaron lattice in opposite to ion beam direction. Moreover, the appearance of very broad line, Figure 2, seems to be indication on the excitation  of the Fr\"ohlich movement of $\sigma$-polaron lattice itself. The presence of Fr\"ohlich sliding of $\sigma$-polaron lattice allows to explain qualitatively the appearance of "hysteresis" in spectral dependences relatively change of the direction of exciting laser wave propagation into opposite direction. It follows from energy law conservation position. Really, moving $\sigma$-polaron lattice possesses by kinetic energy. It means, that it is required lesser energy value to excite the local $\sigma$-polaron vibration mode in given case in correspondence with observation. 

Therefore, CZSNTs incorporated in diamond matrix represent themselves the example of the system, which strongly interact with EM-field. 

The appearance of the lines in IR spectra of another quasi 1D-system - heavily doped carbynes - in the range ($3500 - 5000$) cm$^{-1}$ with peak positions correspondingly at $3729.1 \pm 4$, $3956.2 \pm 10$ cm$^{-1}$, $4409.2 \pm 4$ cm$^{-1}$,  $4720.8 \pm 10$ cm$^{-1}$ in the spectrum of $A$-sample, Figure 3, and at $3892.6 \pm 4$ cm$^{-1}$, $4174.6 \pm 4$ cm$^{-1}$, $4420.3 \pm 4$ cm$^{-1}$ in the spectrum of $B$-sample, Figure 4, seems to be direct indication of the emergence in given samples of longlived coherent state, determined by the space propagation of quantum Rabi oscillations. Really, in the work \cite{Yearchuck_Yerchak_Dovlatova} was established theoretically, that  in the case of strong electron-(spin)-photon interaction, there is the  possibility to observe optical spectra in the
same samples  in usual deterministic regime and in
stochastic regime by the change of registration conditions.  It was concluded immediately from mathematical structure of new  difference-differencial equations for dynamics of spectroscopic transitions for both radio- and optical spectroscopy (instead Bloch equations) for the model, representing itself the 1D-chain of N two-level  equivalent elements coupled by exchange interaction (or its optical analogue for the optical transitions) between themselves and interacting with quantized EM-field and   quantized lattice deformation (phonon) field. The switch to stochastic regime has been achieved by observation  of both ferroelectric
spin wave resonance and  antiferroelectric spin wave resonance \cite{Yearchuck_PL}, \cite{Yearchuck_Yerchak_Dovlatova} by means of  IR-reflection, absorption and transmittance spectroscopy
in the same carbynoid samples. It seems to be undestandable, that coherence will get broken in stochastic regime resulting in disappearance in the spectra of the lines determined by the space propagation of quantum Rabi oscillations. At the same time, if the lines in spectral range  ($3500 - 5000$) cm$^{-1}$ belong to the so called second order transitions, then they  have to be presenting in stochastic regime too, especially if to take into account, that in deterministic regime they are registered with the amplitude of the same order with main lines, see Figure 3. We see from Figure 2 in \cite{Yearchuck_Yerchak_Dovlatova}, that actually the lines in spectral range  ($3500 - 5000$) cm$^{-1}$ disappeared, instead lines the quantum noise is appeared with the average amplitude almost linearly increasing with frequency increase  in the range  ($3500 - 5000$) cm$^{-1}$. In the favour of the interpretation proposed shows also the analysis of the shape of the line at $4409.2 \pm 4$ cm$^{-1}$
in IR-spectrum of $A$-sample, Figure 3, and the line at $4420.3 \pm 4$ cm$^{-1}$ in IR-spectrum of $B$-sample, Figure 4, which are not overlapped with other spectral lines.

It is seen, that given lines   have characteristic asymmetric Dyson shape \cite{Dyson}. Dyson shape of spectral lines is characteristic, for instance, for ESR absorbtion in metals or in low-resistance semiconductors and it is determined by the space dispersion contribution \cite{Erchak_Zaitsev_Stelmakh}, which is appeared  in conductive media to be consequence of phase change of propagating EM-wave and/or spin polarization transfer itself. It was shown in \cite{Erchak_Zaitsev_Stelmakh}, that, for example, in the case of static paramagnetic centers (that is without spin polarization transfer) in conductive media the space dispersion contribution and absorption contribution to resulting ESR responce is 1 to 1 (for the samples, thickness of which is greatly exceeds the  scin-layer depth). It seems to be  understandable, that the propagation of Rabi oscillations will lead  qualitatively to the same picture, that is, to the space dispersion contribution in the IR-absorption, or IR-reflection response, that actually takes place, Figures 3, 4. 

Therefore, experimental detection of Rabi wave packets confirms on the one hand the theory, elaborated in \cite{Dovlatova_A_Yearchuck_D},   \cite{Yerchuck_D_Dovlatova_A} and reviewed in given Section. It was also the first experimental confirmation of Rabi wave phenomenon, predicted in \cite{Slepyan_Yerchak}, on the other hand.  It means also, that semiclassical description of spectroscopic transitions in quasi-1D CZSNTs and in the systems like them cannot be appropriate. It substantially  raises the practical concernment of QED-theory, especially, if to take into account, that the space propagation of quantum Rabi oscillation leads to formation of coherent states with very long life times, that has great significance for a number of practical applications, in particular, by elaboration of logic
quantum systems including quantum computers and quantum communication systems.

The requirement of  one-dimemensionality is substantial for applicability of Slater principle to CZSNTs \cite{Dovlatova_A_Yearchuck_D}, at the same time the Rabi wave phenomenon seems to be general and can be observed in 2D and 3D-systems.   In given Section we reconsider the interpretation of Raman spectra in new 2D carbon material - graphene. Two kinds of modes are observed in Raman spectra of graphene, the so called first-order RS-peak $G$ and so called second-order lines including $2D$-mode \cite{Calizo}, which is attributed to $D$-peak overtone.   $G$-mode is  inerpreted to be zone-center optical modee \cite{Calizo}. So-called $D$- and $G$-bands lie in carbon materials at around (1330-1360) cm$^{-1}$ and 1580 cm$^{-1}$ respectively for visible excitation. The $D$-peak is usually very intense in amorphous carbon samples, while it is
absent in perfect graphitic samples and in graphene. The $D$-peak according to \cite{Ferrari} corresponds to modes associated with transverse optical
 phonons around the edge of the Brillouin zone. In the molecular
picture, it is associated with the breathing mode of the $sp^2$ aromatic rings \cite{Tuinstra}, \cite{A_Ferrari}. It is also very remarkable, that $2D$-mode, which is observed in the range (2660 - 2710) cm$^{-1}$    is
always visible even when the $D$-peak is absent. Given peculiar behavior is interpreted in the literature by  the double
resonance  activation mechanism of the $D$-peak \cite{Thomsen}, which requires the presence of defects
for its initiation. The following explanation was proposed. In a double resonance process, Raman scattering is a four-step process:
(i) a laser induced generation of an electron-hole pair; (ii) electron-phonon scattering with an
exchanged momentum $\vec{q} ~ \vec{K}$; (iii) electron scattering from a defect, whose recoil absorbs the
momentum of the electron-hole pair; (iv) electron-hole recombination. The requirements
of conservation of energy and momentum can only be satisfied by a defect presence. In a
perfect sample, momentum conservation would be violated by the double resonance mechanism, and thus
the $D$-peak is absent. Momentum conservation however is always satisfied in the case of the $2D$-peak, without the need for defect activation, since the process involves two phonons with
opposite momentum vectors.

  Given explanation of absence of $D$-peak in perfect graphitic samples and in graphene seems to be  vague, especially if to take into consideration,  that $D$-peak at 1332 cm$^{-1}$ is 
marvelously observed in perfect diamond single crystals without additional requirement of structural defect presence. The range (2660 - 2710) cm$^{-1}$ of observation of $2D$-peak indicates on rather large deviation from precisely twice frequency value. Therefore, it seems to be reasonable to assign  $D$-peak with  $sp^3$ bond hybridization, then its absence in perfect graphitic samples and in graphene becomes natural explanation. It means in its turn, that $2D$-peak has quite different origin. 

It seems to be essential, that along with $2D$-peak the other lines were detected.
 So in \cite{Calizo}, the first-order RS-peak $G$ was detected in the vicinity of 1580 $cm^{-1}$ and so called second-order lines were observed around 2480, 2700 ($2D$ band), 3250 cm$^{-1}$ by $\lambda_{exc}$ = 488 nm. It is especially interesting, that the intensity of the  second-order $2D$-band was found to be substantially exceeding the intensity of the corresponding first-order line $G$ \cite{Calizo}. It was roughly four times
the intensity of the $G$ peak by $\lambda_{exc}$ = 488 nm. At the same time $2D$-band was strongly suppressed and the lines at 2480,  3250 cm$^{-1}$ were even not detected under ultraviolet (UV) excitation with $\lambda_{exc}$ = 325 nm \cite{Calizo}, that was not explained. Moreover, the $2D$-band under UV excitation shifts to the larger wave numbers and is found near 2825 cm$^{-1}$. At the same time the $G$-peak on all spectra appears at the same position at 1580 cm$^{-1}$. The shift of about 185 cm$^{-1}$ of $2D$-peak from double $D$-peak frequency value seems to be too large to ascribe given line to $D$-peak overtone. 

  It is also seen,  that
 theoretical dependence of $2D$-band position on excitation energy in the framework of the resonant Raman scattering model, presented  in \cite{Calizo} by Figure 5, is far from experimental dependence. At the same time the suggestion, that "the second-order" lines are spectral mapping  of revival part of Rabi wave packet, corresponding to  the first-order line $G$, is agreeing  with experimental data in \cite{Calizo} very well, see Figure 5 in our paper. The dependence, presented in Figure 5,  was built from experimental data in \cite{Calizo} by choosing instead the energy the electrical component of oscillating excitation field to be $x$-coordinate.  It is seen,    that given experimental dependence is linear in full correspondence with expected for the center of Fourier transform of revival part of Rabi wave packets. It is analogue of aforesaid well known in radiospectroscopy linear dependence of Rabi oscillation frequency on magnetic  component of oscillating excitation field, see, for instance, \cite{Yerchak}, \cite{Yerchak_Stelmakh}. 

Strong suppression of the "second order lines" including vanishing the lines at 2480,  3250 cm$^{-1}$ under UV excitation with $\lambda_{exc}$ = 325 nm can be easily explained by essentially more wide spectrum of nonresonance phonons, taking part in relaxation processes under UV excitation in comparison with the spectrum of nonresonance phonons under  excitation with visible light. It is well known, that it leads to strong supression of quantum Rabi oscillation \cite{Zhu} and the peculiarity,  consisting in strong suppression of the "second order lines"  becomes therefore the natural explanation.

Thus,  the  experimental results presented  \cite{Calizo} are in fact experimental evidence for the formation and propagation of Rabi wave packets in graphene. It seems to be evident, that   Rabi wave packets  can also be identified in free standing 2D-NTs by the production technology improvement.

Similar explanation can be proposed for   transitions in heavily boron-doped
diamond in the region of 1400-2800 $cm^{-1}$ \cite{{Vlasov}}. It means, that  the theory of Rabi wave formation, presented in \cite{Slepyan_Yerchak}   for quasi-1D-systems can be  developed and generalized for 2D- and 3D-systems.   

 \section{Formation of Longlived Coherent Resonance Phonon System by Spectroscopic Transitions}

It has been established in recent work \cite{Dovlatova_Yerchuck}, that by strong dipole-photon and dipole-phonon coupling the   formation of longlived coherent system of the resonance phonons takes place,  and relaxation processes acquire pure quantum character. It is determined by the appearence of coherent emission process of EM-field energy, for which the resonance phonon system is responsible. Emission process is  accompanying by phonon Rabi quantum oscillations,  which can be time-shared from photon quantum Rabi  oscillations, accompanying coherent absorption process of EM-field energy. For the case of radiospectroscopy it corresponds to the possibility of the simultaneous observation  along with (para)magntic spin resonance the acoustic spin resonance.  

Let us reproduce the proof of given conclusion for the convenience of readers in details.

Given work is development of the 
 work \cite{Yearchuck_Yerchak_Dovlatova}, where the system of  difference-differencial equations for dynamics of spectroscopic transitions for both radio- and optical spectroscopy for the model, representing itself the 1D-chain of N two-level  equivalent elements coupled by exchange interaction (or its optical analogue for the optical transitions) between themselves and interacting with quantized EM-field and   quantized phonon field its optical  has recently been  derived. Naturally the equations are true for any 3D system of paramagnetic centers (PC) or optical centers  by the absence of exchange interaction. In given case the model presented  differs from Tavis-Cummings model \cite{Tavis} by inclusion into consideration of quantized phonon system, describing the relaxation processes from quantum fied theory position.  Seven equations for the seven operator variables, describing joint system \{field + matter\} can be presented in matrix form by three matrix equations. They are the following

\begin{equation}
\label{eq1}
\begin{split}
\raisetag{40pt}
\frac{\partial}{\partial t} 
\left[\begin{array}{*{20}c}
{\hat\sigma^-_l}  \\
 \\
{\hat\sigma^+_l}  \\
\\
{\hat\sigma^z_l} 
\end{array} 
\right] = 2\left\|g\right\|\left[\begin{array}{*{20}c}{\hat F^-_l}  \\
 \\
{\hat F^+_l}  \\
\\
{\hat F^z_l} 
\end{array} 
\right] +  ||\hat{R}^{(\lambda)}_{\vec{q}l}||, 
\end{split}
\end{equation}

\begin{equation}
\label{eq2}
\begin{split}
\raisetag{40pt}
&\frac{\partial}{\partial t} 
\left[\begin{array}{*{20}c}
 {\hat{a}_{\vec k^{}}} \\
 \\
 {\hat{a}_{\vec k^{}}^+} \\
\end{array} 
\right] = -i \omega_{\vec k^{}} ||\sigma_P^z|| \left[\begin{array}{*{20}c}
 {\hat{a}_{\vec k^{}}} \\
 \\ 
 {\hat{a}_{\vec k^{}}^+} \\
\end{array} 
\right] \\
\\
& + \frac{i}{\hbar}
\left[\begin{array}{*{20}c}
{-\sum\limits_{l = 1}^N (\hat\sigma_l^{+} + \hat\sigma_l^{-}) v_{l \vec k}^*} \\
\\
{\sum\limits_{l = 1}^N (\hat\sigma_l^{+} + \hat\sigma_l^{-}) v_{l \vec k}} \\
\end{array} \right],
\end{split}
\end{equation}

\begin{equation}
\label{eq3}
\begin{split}
\raisetag{40pt}
\frac{\partial}{\partial t} 
\left[
\begin{array}{*{20}c}
 {\hat{b}_{\vec k^{}}} \\
 \\
 {\hat{b}_{\vec q^{}}^+} \\
\end{array} 
\right] = -i \omega_{\vec q^{}} ||\sigma_P^z|| \left[\begin{array}{*{20}c}
 {\hat{b}_{\vec q^{}}} \\
 \\ 
 {\hat{b}_{\vec q^{}}^+} \\
\end{array} 
\right] 
 + \frac{i}{\hbar}
\left[
\begin{array}{*{20}c}
{-\sum\limits_{l = 1}^N  \hat\sigma_l^{z} \lambda_{\vec q l}} \\
\\
{\sum\limits_{l = 1}^N \hat\sigma_l^{z} \lambda_{\vec q l}} \\
\end{array} \right],
\end{split}
\end{equation}
where
\begin{equation}
\label{eq4}
\begin{split}
\raisetag{40pt}
\left[\begin{array}{*{20}c}
{\hat\sigma^-_l}  \\
 \\
{\hat\sigma^+_l}  \\
\\
{\hat\sigma^z_l} 
\end{array} 
\right] = \hat{\vec{\sigma}}_l =  \hat\sigma^-_l  \vec e_ +  +  \hat\sigma^+_l \vec e_ - +  \hat\sigma^z_l\vec e_z
\end{split}
\end{equation} is  vector-operator of spectroscopic transitions for $l$th chain unit, $l = \overline{2,N-1}$ \cite{Yearchuck_Yerchak_Dovlatova}.
Its components, that is,  the operators 
\begin{equation}
\label{eq6a}
{\hat\sigma_v}^{jm} \equiv {\left|j_v \right\rangle} {\left\langle m_v \right|} 
\end{equation} are set up in correspondence to the states ${\left|j_v \right\rangle}$,${\left\langle m_v \right|}$, where $v = \overline{1,N}$, 
$j = \alpha, \beta$, $m = \alpha, \beta $. For instance, the relationships for commutation rules are
\begin{equation}
\label{eq9a}
[\hat {\sigma}_v^{lm}, \hat {\sigma}_v^{pq}] = \hat {\sigma }_v^{lq} \delta_{mp} - \hat {\sigma }_v^{pm}\delta_{ql}. 
\end{equation} 
Further
\begin{equation}
\label{eq5}
\begin{split}
\raisetag{40pt}
\left[\begin{array}{*{20}c}
{\hat F^-_l}  \\
 \\
{\hat F^+_l}  \\
\\
{\hat F^z_l} 
\end{array} 
\right] = \hat {\vec F} =  \left[ {\hat {\vec {\sigma}}_l \otimes \hat {\vec {\mathcal{G}}}_{l - 1,l + 1}} \right],
\end{split}
\end{equation}
where vector operators $\hat {\vec {\mathcal{G}}}_{l - 1,l + 1}$,  $l = \overline{2,N-1}$, are given by the expressions
\begin{equation}
\label{eq6}
\hat {\vec {\mathcal{G}}}_{l - 1,l + 1} = \hat {\mathcal{G}}_{l - 1,l + 1}^-  \vec e_ +  + \hat {\mathcal{G}}_{l - 1,l + 1}^ +  \vec e_ - + \hat {\mathcal{G}}_{l - 1,l + 1}^z \vec e_z,
\end{equation}
in which 
\begin{subequations}
\label{eq7}
\begin{gather}
\hat {\mathcal{G}}_{l - 1,l + 1}^-  = -\frac{1 }{\hbar} \sum\limits_{\vec k}\hat{f}_{l \vec k} - \frac{J }{\hbar }(\hat\sigma _{l + 1}^- + \hat\sigma _{l - 1}^-) , \\
\hat {\mathcal{G}}_{l - 1,l + 1}^+ = -\frac{1}{\hbar} \sum\limits_{\vec k}\hat{f}_{l \vec k} - \frac{J}{\hbar }(\hat\sigma _{l + 1}^+ + \hat\sigma _{l - 1}^ + ), \\
\hat {\mathcal{G}}_{l - 1,l + 1}^z = - \omega_{l} - \frac{J}{\hbar }(\hat\sigma _{l + 1}^z + \hat\sigma _{l - 1}^z ).
\end{gather}
\end{subequations}
Here operator $\hat{f}_{l \vec k}$ is
 \begin{equation}
\label{eq8}
\hat{f}_{l \vec k} = v_{l \vec k} \hat{a}_{\vec k} + \hat{a}_{\vec k}^{+} {v^*}_{l \vec k}.
\end{equation}
In relations (\ref{eq7}) $J$ is the exchange interaction constant in the case of magnetic resonance transitions or its optical
analogue in the case of optical transitions, the function $v_{l \vec k}$ in (\ref{eq8}) is
\begin{equation}
\label{eq9}
 v_{l \vec k} = - \frac{1}{\hbar} p_l^{jm} (\vec e_{\vec k} \cdot \vec e_{\vec P_{l}}) \mathfrak{E}_{\vec k} e^{ - i \omega_{\vec k}t + i \vec k \vec r},
\end{equation}
where $p_l^{jm}$ is matrix element of operator of magnetic (electric) dipole moment $\vec P_{l}$ of $\textit{l-th}$ chain unit between the states $\left| {j_{l}} \right\rangle$ and 
$\left| m_{l} \right\rangle$ with $j \in \{\alpha, \beta\}$,  $m \in \{\alpha, \beta\}$, $j \neq m$, $\vec e_{\vec k}$ is unit polarisation vector, $\vec e_{\vec P_{l}}$ is unit vector along $\vec P_{l}$-direction, s$\mathfrak{E}_{\vec k}$ is the quantity, which has the dimension of magnetic (electric) field strength, $\vec k$ is quantized EM-field wave vector, the components of which get a discrete set of values, $\omega_{\vec k}$ is the frequency, corresponding to ${\vec k}$th mode of EM-field,  $\hat{a}^+_{\vec k}$ and $\hat{a}_{\vec k}$ are  EM-field  creation and  annihilation operators correspondingly. In the suggestion, that the contribution of spontaneous emission is relatively small, we will have $p_{l}^{jm} = p_{l}^{mj} \equiv p_{l} $, where $j \in \{\alpha, \beta\}$,  $m \in \{\alpha, \beta\}$, $j \neq m$. Further matrix $||\hat{R}^{(\lambda)}_{\vec{q}l}||$ is
\begin{equation}
\label{eq10}
\begin{split}
\raisetag{40pt}
||\hat{R}^{(\lambda)}_{\vec{q}l}|| = 
\frac{1}{i\hbar} \left[\begin{array}{*{20}c}{ 2 \hat{B}^{(\lambda)}_{\vec{q}l} \hat\sigma^-_l}  \\
 \\
{ -2 \hat{B}^{(\lambda)}_{\vec{q}l} \hat\sigma^+_l}  \\
\\
{0} \end{array} 
\right] 
\end{split}
\end{equation}
Here $\hat{B}^{(\lambda)}_{\vec{q}l}$ is
\begin{equation}
\label{eq10a}
\hat{B}^{(\lambda)}_{\vec{q}l} = \sum\limits_{\vec{q}}\lambda_{\vec{q}l} (\hat{b}^{+}_{\vec{q}} + \hat{b}_{\vec{q}}),
\end{equation} 
 $\hat{b}_{\vec q}^+$ ($\hat{b}_{\vec q}$) is the creation
(annihilation) operator of the phonon with impulse ${\vec q}$ and with
energy $\hbar \omega_{\vec q} $, $\lambda_{\vec q l}$ is electron-phonon coupling constant. In equations (\ref{eq2}) and (\ref{eq3})
$\|\sigma_P^z\| $ is Pauli $z$-matrix,  $\left\| g \right\|$ in equation (\ref{eq1}) is diagonal matrix,
numerical values of its elements are dependent on the basis choice.
It is at appropriate basis
\begin{equation}
\label{eq11}
\left\| g \right\| = 
\left[
\begin{array}{*{20}c}
 {1} & {0} & {0} \\
 {0} & {1} & {0} \\
 {0} & {0} & {1} \\
\end{array} 
\right]. 
\end{equation} 
 
Right hand side expression  in (\ref{eq5}) is vector product of vector operators.   It can be 
calculated by using of known expression (\ref{eq12}) with additional coefficient ${\frac{1}{2}}$ the only, which is appeared, since
\begin{equation}
\label{eq12}
\left[ {\hat {\vec {\sigma}} _l \otimes \hat {\vec {\mathcal{G}}}_{l - 1,l + 1} } \right] = \frac{1}{2} \left| {\begin{array}{*{20}c}
 {\vec e_- \times \vec e_z} & {\hat{\sigma}_l^-} & {\hat {\mathcal{G}}_{\,\,l - 1,l + 1}^-} \\
 {\vec e_z \times \vec e_+} & {\hat{\sigma}_l^+} & {\hat {\mathcal{G}}_{\,\,l - 1,l + 1}^+} \\
 {\vec e_+ \times \vec e_-} & {\hat{\sigma}_l^z} & {\hat {\mathcal{G}}_{\,\,l - 1,l + 1}^z} \\
\end{array}} \right|',
\end{equation}
the products of two components of two vector operators are replaced by anticommutators of corresponding 
components. Given detail is mapped by symbol $\otimes$ in (\ref{eq5}) and by symbol $'$ in determinant (\ref{eq12}). 
The equation, which is given by (\ref{eq1}) is  QED-generalization of 
 semiclassical Landau-Lifshitz (L-L) equation for dynamics of spectroscopic transitions in a chain of exchange coupled centers derived in \cite{Yerchak_Yearchuck} and solved analytically in \cite{Yearchuck_Yerchak_Dovlatova}.
 In comparison with semiclassical description, where the description of dynamics of spectroscopic 
transitions is exhausted by one vector L-L equation equation, in the case of completely quantum consideration L-L type equation describes the only one subsystem of three-part-system, which consist of EM-field, dipole moments' (magnetic or electric) matter subsystem and phonon subsystem. It was concluded in \cite{Yearchuck_Yerchak_Dovlatova}, that the presence of additional equations for description of transition dynamics by QED model in comparison with semiclassical model  leads  to a number of  new effects, which can be predicted the only by QED consideration of resonance transition phenomena. One of new effect was described in 
\cite{Yearchuck_Yerchak_Dovlatova}, starting the only from the mathematical structure of the equations. It was argued, that the equations (\ref{eq1}), (\ref{eq2}) represent
themselves vector-operator difference-differential generalization of the system, which belongs to well known family of equation systems - Volterra model systems, widely used in biological tasks of population dynamics studies, which in its turn is generalization of Verhulst equation.  In other words, it was predicted, for instance, that by some parameters in two-sybsystem Volterra model the stochastic component in solution will be appeared. Given prediction has aforesaid experimental confirmation by the study of optical properties in carbynes \cite{Yearchuck_PL} indicating, that in given material strong electron-photon interaction is realized, which allows to explain the possibility to observe the stationary IR-reflection or absorption sspectra both in usual and in stochastic regime. 

The terms like to right hand side terms in (\ref{eq3})  were used in so called "spin-boson" Hamiltonian \cite{Leggett} and in  so called "independent boson model" \cite{Mahan}. Given models were used to study phonon effects in a single quantum dot within a microcavity \cite{Heitz}, \cite{Tuerck}, \cite{Besombes},   \cite{WilsonRae}, \cite{Zhu}. So, it has been shown in \cite{WilsonRae}, \cite{Zhu}, that the presence of the term in Hamiltonian \cite{Yearchuck_Yerchak_Dovlatova}
\begin{equation}
\label{eq13}
\mathcal{\hat H}^{CPh} =\sum\limits_{j=1}^N \sum\limits_{\vec q}  \lambda_{\vec q} (\hat{b}_{\vec q}^{ +} +\hat{b}_{\vec q})\hat{\sigma}^z_j,
\end{equation} which coincides with corresponding term in Hamiltonian in \cite{WilsonRae}, \cite{Zhu} at $N = 1$ [contribution of given term to  the equations for spectroscopic transitions  is  $\pm {\sum\limits_{l = 1}^N \hat\sigma_l^{z} \lambda_{\vec q}}$, see equation (\ref{eq3}), (note that the equations for spectroscopic transitions were not derived in above cited works \cite{Heitz}, \cite{Tuerck}, \cite{Besombes},   \cite{WilsonRae}, \cite{Zhu})] leads the only to exponential decrease  of the magnitude of quantum Rabi oscillations with increase of electron-phonon coupling strength and even to their supression at relatively strong electron-phonon coupling.  

However, by strong electron-photon coupling and strong electron-phonon coupling quite other picture of quantum relaxation processes becomes to be possible. Really,
if to define the 
 wave function  of the chain system, interacting
with quantized EM-field  and with quantized lattice vibration field, to be
vector of the state in Hilbert space over quaternion ring,  that is 
quaternion function of quaternion
argument, then like to \cite{Yearchuck_Yerchak_Dovlatova} can be shown, that the equations (\ref{eq1}) to (\ref{eq3}) are Lorentz invariant and the transfer to observables can be realized. In particular, taking into account, that quaternion vector of the state is proportional to spin, the   Hamiltonian, given by  (\ref{eq13}) describes in fact the interaction of phonon field  with $z$- component $S^z$ of the spin of matter subsystem. It seems to be reasonable to take into consideration the  interaction of phonon field with  $S^+$- and $S^-$ components of the spin of matter subsystem. Therefore, in \cite{Dovlatova_Yerchuck}  in a natural way  the following structure of  Hamiltonian was proposed 
 \begin{equation}
\label{eq14}
\mathcal{\hat H} = \mathcal{\hat H}^C + \mathcal{\hat H}^F + \mathcal{\hat H}^{C F} + \mathcal{\hat H}^{Ph} + \mathcal{\hat H}^{CPh} ,
\end{equation} 
where 
${\mathcal{\hat H}^C}$ is chain Hamiltonian by the absence of the interaction with EM-field, ${\mathcal{\hat H}^F}$ is field Hamiltonian, ${\mathcal{\hat H}^{C F}}$ is Hamiltonian, describing the interaction between quantized EM-field and atomic chain.
 Hamiltonian ${\mathcal{\hat H}^C}$ is
\begin{equation}
\label{eq15}
{\mathcal{\hat H}^C} = {\mathcal{\hat H}^0} + {\mathcal{\hat H}^J}, 
\end{equation} 
where ${\mathcal{\hat H}^0}$ is chain Hamiltonian in the absence of the interaction between structural elementary units of the chain.
  ${\mathcal{\hat H}^0}$ is given by the expression
\begin{equation}
\label{eq16}
\mathcal{\hat H}^0 = \sum\limits_{v=1}^N \sum\limits_m E_{mv}{\left|m_v \right\rangle} {\left\langle m_v \right|}.
\end{equation}
Here $m = \alpha, \beta$, $E_{mv}$ are eigenvalues of ${\mathcal{\hat H}^0}$, which correspond to the states ${\left|m_v \right\rangle}$ of $v\textit{th}$ chain unit.                   
Hamiltonian ${\mathcal{\hat H}^J}$ is
\begin{equation}
\label{eq17}
\begin{split}
\raisetag{40pt}
 \mathcal{\hat H}^J = 
 \sum\limits_{n = 1}^N [J_{_{E}} (\hat {\sigma} _n^ + \hat {\sigma} _{n + 1}^- + \hat {\sigma} _n^ -  \hat {\sigma} _{n + 1}^ + + \frac{1}{2}\hat {\sigma} _n^z \hat {\sigma} _{n + 1}^z ) + H.c.].
\end{split}
\end{equation}
It is suggested 
in the model, that $\left| {\alpha _n } \right\rangle $ and $\left| {\beta _n } \right\rangle $ are eigenstates, producing the full set for each of $N$ elements. It is evident, that given assumption can be realized strictly the 
only by the absence of the interaction between the elements. At the same 
time proposed model will rather well describe the real case, if the 
interaction energy of adjacent elements is much less of the energy of the 
splitting $\hbar \omega _0 =\mathcal{E}_\beta -\mathcal{E}_\alpha$ between the energy levels, 
corresponding to the states $\left|\alpha_n\right\rangle$ and $\left|\beta_n\right\rangle$. The case considered includes in fact all known 
experimental situations.
  Hamiltonian $\mathcal{\hat H}^{C F}$ of interaction of quantized EM-field with atomic chains was also  represented in the set of variables, which includes the components of spectroscopic transition vector operator $\hat {\vec {\sigma }}_v$. It is in suggestion of dipole approximation  and  by  fixed polarization of field components  the following 
\begin{equation}
\label{eq18}
\begin{split}
\raisetag{40pt}
\mathcal{\hat H}^{C F} = -\sum\limits_{j = 1}^n \sum\limits_{l \neq m} \sum\limits_{m} \sum\limits_{\vec k} [ p_j^{lm} \hat {\sigma}_j^{lm} (\vec e_{\vec k} \vec e_ {\vec P_j}) \mathfrak{E}_{\vec k} \hat{a}_{\vec k} \times \\
e^{ - i \omega_{\vec k} t+ i \vec k \vec r} + H.c. ], 
\end{split}
\end{equation}
where $p_j^{lm}$ is matrix element of operator of magnetic (electric) dipole moment $\vec P_j$ of $\textit{j-th}$ chain unit between the states $\left| {l_j} \right\rangle$ and 
$\left| {m_j} \right\rangle$ with $l_j = \alpha_j, \beta_j, m_j = \alpha_j, \beta_j$, $\vec e_{\vec k}$ is unit polarization vector, $\vec e_{\vec P_j}$ is unit vector along $\vec P_j$-direction, $\mathfrak{E}_{\vec k}$ is the quantity, which has the dimension of  magnetic (electric) field strength, $\vec k$ is wave vector, $\hat{a}_{\vec k}$ is field annihilation operator. In the suggestion, that the contribution of spontaneous emission is relatively small,  $p_j^{lm} = p_j^{ml} \equiv p_j $, where $l = \alpha, \beta, m = \alpha, \beta$. Then the function 
\begin{equation}
\label{eq19}
 q_{j \vec k} = - \frac{1}{\hbar} p_j (\vec e_{\vec k} \cdot \vec e_{\vec P_j}) \mathfrak{E}_{\vec k} e^{ - i \omega_{\vec k} t+ i \vec k \vec r}
\end{equation}
was defined, and 
 the expression (\ref{eq18}) was rewritten in the form
\begin{equation}
\label{eq20}
\mathcal{\hat H}^{C F} = \sum\limits_{v = 1}^n \sum\limits_{\vec k} [q_{j \vec k} (\hat {\sigma}_j^- + \hat {\sigma}_j^+) \hat{a}_{\vec k} + (\hat {\sigma}_j^- + \hat {\sigma}_j^+) \hat{a}_{\vec k}^{ +} {q^*}_{j \vec k}],
\end{equation}
where $\hat{a}_{\vec k}^{ +}$ is EM-field creation operator, $\hat{a}_{\vec k}^{ }$ is EM-field annihilation operator, superscript $^*$ in ${q^*}_{j \vec k}$ means complex conjugation.
 Field Hamiltonians have usual form
\begin{equation}
\label{eq21}
\mathcal{\hat H}^{F} = \sum\limits_{\vec k} \hbar \omega_{\vec k} (\hat{a}_{\vec k}^{ +} \hat{a}_{\vec k} + \frac{1}{2})
\end{equation}
for EM-field and
\begin{equation}
\label{eq22}
\mathcal{\hat H}^{Ph} = \sum\limits_{\vec q} \hbar \omega_{\vec q} (\hat{b}_{\vec q}^{ +} \hat{b}_{\vec q} + \frac{1}{2})
\end{equation} for phonon field.
 Hamiltonian $\mathcal{\hat H}^{CPh}$ is
\begin{equation}
\label{eq23}
\mathcal{\hat H}^{CPh} = \mathcal{\hat H}^{CPh}_z + \mathcal{\hat H}^{CPh}_\pm, 
\end{equation} 
where $\mathcal{\hat H}^{CPh}_z$ is determined by the expression

\begin{equation}
\label{eq24}
\mathcal{\hat H}^{CPh}_z =\sum\limits_{j=1}^N \sum\limits_{\vec q} (\lambda^z_{\vec q} \hat{b}_{\vec q} + (\lambda^z_{\vec q})^*\hat{b}_{\vec q}^{+})\hat{\sigma}^z_j.
\end{equation}
 Hamiltonian $\mathcal{\hat H}^{CPh}_\pm$ was represented in the following form
\begin{equation}
\label{eq25}
\mathcal{\hat H}^{CPh}_\pm = \sum\limits_{j=1}^N \sum\limits_{\vec q}\lambda^\pm_{\vec q} (\hat{\sigma}^-_j + \hat{\sigma}^+_j) \hat{b}_{\vec q} + (\lambda^\pm_{\vec q})^* (\hat{\sigma}^-_j + \hat{\sigma}^+_j)\hat{b}_{\vec q}^{+}.
\end{equation}
Here  $\lambda^z_{\vec q}$ and $\lambda^\pm_{\vec q}$ are  electron-phonon coupling constants, which characterisire correspondingly the interaction with $z$- component $S^z_j$ and  with  $S^+$- and $S^-_j$ components of the spin of jth chain unit. It seems to be understandable, that they can be different in general case. Moreover, in order to take into account the interaction with both equilibrium and nonequilibrium phonons  both the electron-phonon coupling constants have to be complex numbers, that takes proper account by expressions (\ref{eq24}), (\ref{eq25}). 
It has been shown, that 
ihe equations of the motion for spectroscopic transition operators $\hat {\vec {\sigma }}_l$,  for quantized 
 EM-field operators $\hat{a}_{\vec k}$, $\hat{a}_{\vec k}^{ +}$ and for phonon field operators $\hat{b}_{\vec q}$, $\hat{b}_{\vec q}^{ +}$  are the following. Instead equation (\ref{eq1}) we have
\begin{equation}
\label{eq26}
\begin{split}
\raisetag{40pt}\frac{\partial}{\partial t} \left[\begin{array}{*{20}c}
{\hat\sigma^-_l}  \\
 \\
{\hat\sigma^+_l}  \\
\\
{\hat\sigma^z_l} 
\end{array} 
\right] = 2 \left\| g \right\| \left[\begin{array}{*{20}c}
{\hat F^-_l}  \\
 \\
{\hat F^+_l}  \\
\\
{\hat F^z_l} 
\end{array} 
\right] + ||\hat{R}^{(\lambda^z)}_{\vec{q}l}|| + ||\hat{R}^{(\lambda^\pm)}_{\vec{q}l}||, 
\end{split}
\end{equation}
where matrix $||\hat{R}^{(\lambda^z)}_{\vec{q}l}||$ is
\begin{equation}
\label{eq27}
\begin{split}
\raisetag{40pt}
||\hat{R}^{(\lambda^z)}_{\vec{q}l}|| = 
\frac{1}{i\hbar} \left[\begin{array}{*{20}c}{ 2 \hat{B}^{(\lambda^z)}_{\vec{q}l} \hat\sigma^-_l}  \\
 \\
{ -2 \hat{B}^{(\lambda^z)}_{\vec{q}l} \hat\sigma^+_l}  \\
\\
{0} \end{array} 
\right] 
\end{split}
\end{equation}
with  $\hat{B}^{(\lambda^z)}_{\vec{q}l}$, which is given by
\begin{equation}
\label{eq28}
\hat{B}^{(\lambda^z)}_{\vec{q}l} = \sum\limits_{\vec{q}}[(\lambda^z_{\vec{q}l})^* \hat{b}^{+}_{\vec{q}} + \lambda^z_{\vec{q}l} \hat{b}_{\vec{q}}].\end{equation} 
Matrix $||\hat{R}^{(\lambda^\pm)}_{\vec{q}l}||$ is
\begin{equation}
\label{eq29}
\begin{split}
\raisetag{40pt}
||\hat{R}^{(\lambda^z)}_{\vec{q}l}|| = 
\frac{1}{i\hbar} \left[\begin{array}{*{20}c}{ -\hat{B}^{(\lambda^\pm)}_{\vec{q}l} \hat\sigma^z_l}  \\
 \\
{  \hat{B}^{(\lambda^\pm)}_{\vec{q}l} \hat\sigma^z_l}  \\
\\
{\hat{B}^{(\lambda^\pm)}_{\vec{q}l} (\hat\sigma^+_l - \hat\sigma^-_l)} \end{array} 
\right],  
\end{split}
\end{equation}
where $\hat{B}^{(\lambda^\pm)}_{\vec{q}l}$ is
\begin{equation}
\label{eq30}
\hat{B}^{(\lambda^\pm)}_{\vec{q}l} = \sum\limits_{\vec{q}}[(\lambda^\pm_{\vec{q}l})^* \hat{b}^{+}_{\vec{q}} + \lambda^\pm_{\vec{q}l} \hat{b}_{\vec{q}}].\end{equation} 
The equation (\ref{eq2}) remains without changes. The equation (\ref{eq3}) is
\begin{equation}
\label{eq30a}
\begin{split}
\raisetag{40pt}
&\frac{\partial}{\partial t} 
\left[
\begin{array}{*{20}c}
 {\hat{b}_{\vec k^{}}} \\
 \\
 {\hat{b}_{\vec q^{}}^+} \\
\end{array} 
\right] = -i \omega_{\vec q^{}} ||\sigma_P^z|| \left[\begin{array}{*{20}c}
 {\hat{b}_{\vec q^{}}} \\
 \\ 
 {\hat{b}_{\vec q^{}}^+} \\
\end{array} 
\right] 
 + \\
&\frac{i}{\hbar}
\left[
\begin{array}{*{20}c}
{-\sum\limits_{l = 1}^N \{\lambda^z_{\vec q l} \hat\sigma_l^{z} + \lambda^{\pm}_{\vec q l} (\hat\sigma_l^{+} + \hat\sigma_l^{-})\}} \\
\\
{\sum\limits_{l = 1}^N \{\lambda^z_{\vec q l} \hat\sigma_l^{z} + \lambda^{\pm}_{\vec q l} (\hat\sigma_l^{+} + \hat\sigma_l^{-})\}} \\
\end{array} \right].
\end{split}
\end{equation}

 Thus, QFT model for  dynamics of spectroscopic transitions in 1D multiqubit  exchange coupled  system  is generalized by taking into account the earlier proof \cite{Yearchuck_Yerchak_Dovlatova}, that spin vector is  quaternion vector of the state of any quantum systen in Hilbert space defined  over quaternion ring and consequently all the spin components has to be taken into account. New quantum  phenomenon was predicted in \cite{Dovlatova_Yerchuck}. The prediction results from the structure of  the equations derived and it consists in the following. The coherent system of the resonance phonons, that is,  the phonons with the energy, equaled to resonance photon  energy can be formed by resonance, that can lead to appearance along with   Rabi oscillations determined by spin (electron)-photon coupling with the frequency $\Omega^{RF}$ of Rabi oscillations determined by spin (electron)-phonon coupling with the frequency $\Omega^{RPh}$. In other words,  QFT model predicts the oscillation character of quantum relaxation, that is quite different character in comparison with phenomenological and semiclassical Bloch models. Moreover, if $\mid\lambda^{\pm}_{\vec q l}\mid < g$ the second Rabi oscillation process will be observed by stationary state of two subsystems \{EM-fied + magnetic (electric) dipoles\}, that is, it will be registered in quadrature with the first Rabi oscillation process.  It can be experimentally detected even by stationary spectroscopy methods.

\section{Conclusions} 

Brief review of the theoretical and experimental results, based mainly on the works of authors, in the application of quantum field theory  to the study of carbon low-dimensional systems - quasi-1D carbon nanotubes, carbynes and graphene with  emphasis on formation of longlived coherent states of joint photon-electron and joint resonance phonon-electron systems of given materials is presented. Two new ways for  longlived coherent state formation are considered in 
present work,  leading to essentially  more long lifetimes in comparison with known methods of formation of individual coherent states, realised on point centers in crystals.  The first way is the
longlived coherent state formation in  result of Rabi-wave packets' propagation in the materials with strong electron-photon interaction. Given mechanism was confirmed experimentally on examples of a number carbon low-dimensional materials -  quasi-1D carbon nanotubes, carbynes,  graphene. Naturally, the range of suitable materials for given aim is much wider.  

Main difference of given states in comparison with individual coherent states, realised on point centers, that they are collective coherent states.

On the other hand the appearance of additional lines, associated with the same RS-mode, means, that  strong electron-photon coupling takes place in quasi-1D CZSNTs, carbynes and  graphene by interaction with EM-field and quantum nature of EM-field has to be taken into account in any experiments and practical application of quasi-1D CZSNTs  carbynes and  graphene with participation of EM-field. By strong electron-photon coupling all optical spectra, in particular, Raman spectra seem to be registered (in spite of the usage of stationary measurement technique)  in nonequilibrium coherent state, which is the consequence of Rabi wave packets' formation and propagation.

QFT model for  dynamics of spectroscopic transitions in 1D multiqubit  exchange coupled  system  is generalized by taking into account the earlier proof \cite{Yearchuck_Yerchak_Dovlatova}, that spin vector is  quaternion vector of the state of any quantum systen in Hilbert space defined  over quaternion ring and consequently all the spin components has to be taken into account. New quantum  phenomenon is predicted. The prediction results from the structure of  the equations derived and it consists in the following. The coherent system of the resonance phonons, that is,  the phonons with the energy, equaled to resonance photon  energy can be formed by resonance, that can lead to appearance along with   Rabi oscillations determined by spin (electron)-photon coupling with the frequency $\Omega^{RF}$ of Rabi oscillations determined by spin (electron)-phonon coupling with the frequency $\Omega^{RPh}$. In other words QFT model predicts the oscillation character of quantum relaxation, that is quite different character in comparison with phenomenological and semiclassical Bloch models.
  Moreover if  absolute value of  electron-phonon coupling constant $|\lambda^{\pm}_{\vec q l}|$, which characterises  the interaction with    $S^+$- and $S^-_j$ components of the spin of jth chain unit is less, than electron(spin)-photon coupling constant $g$,  the model predicts, that the second quantum Rabi oscillation process will be observed by stationary state of joint two subsystems \{EM-field + magnetic (electric) dipoles\}, and it  will be registered in quadrature with the first quantum Rabi oscillation process. The second quantum Rabi oscillation process is governed by  the formation of the coherent system of the resonance phonons. Therefore along with absorption process of EM-field energy the coherent emission process takes place. Both the quantum Rabi oscillation processes can be time-shared. For the case of radiospectroscopy it corresponds to the possibility of the simultaneous observation  along with (para)magnetic spin resonance the acoustic spin resonance. The second (acoustic) quantum Rabi oscillation process can be detected even  by stationary spectroscopy methods.

New quantum  phenomenon  predicted opens  the second way of the formation of the coherent states of joint system \{magnetic (electric) dipoles + resonance phonons\} in the matter. 
Both the  phenomena of  the formation of the coherent states in the matter  can find a   number of practical applications, in particular they can be used by elaboration of various logic quantum systems including quantum computers and quantum communication systems, in which quantized EM-field and/or quantized acoustic field will be working components. The same conclusion is concerned the elaboration of various optoelectronic and spintronic devices.

\end{document}